\documentclass [] {aa} 

\usepackage{txfonts,graphicx,lscape,natbib}
\bibpunct{(}{)}{;}{a}{}{,} 

\begin{document}

\title{Kinematics of planet-host stars and their relation with dynamical streams in the solar neighbourhood
\thanks{}}

\author{A.~Ecuvillon\inst{1}, G.~Israelian\inst{1},  F.~ Pont\inst{2}, N. C.~ Santos\inst{2,3,4} \and M.~Mayor\inst{2}}

\offprints{\email{aecuvill@ll.iac.es}}

\institute{Instituto de Astrof\'{\i}sica de Canarias, E-38200 La Laguna, Tenerife, Spain 
\and Observatoire de Gen\`eve, 51 ch.  des  Maillettes, CH--1290 Sauverny, Switzerland
\and Centro de 
Astronomia e Astrofisica de Universidade de Lisboa, Observatorio Astronomico de Lisboa, Tapada de Ajuda, 1349-018 Lisboa, 
Portugal
\and Centro de Geofisica de \'Evora, Rua Roma\~o Ramalho 59, 7000 \'Evora, Portugal 
}
\date{Received 22 June 2006/ Accepted 28 August 2006} 

\titlerunning{Kinematics of planet-host stars} 
\authorrunning{A. Ecuvillon et al.}

\abstract{We present a detailed study on the kinematics of metal-rich stars with and without planets, 
and their relation with the Hyades, Sirius and Hercules dynamical streams  in the solar neighbourhood. 
Accurate kinematics have been derived for all the stars belonging to the CORALIE planet search survey. 
We used precise radial velocity measurements and CCF parameters from the CORALIE database, and parallaxes, 
photometry and proper motions from the HIPPARCOS and Tycho-2 catalogues. The location of stars with planets 
in the thin or thick discs has been analysed using both kinematic and chemical constraints. We compare 
the kinematic behaviour of known planet-host stars to the remaining targets belonging to the volume-limited 
sample, in particular to its metal-rich population. The high average metallicity of the Hyades stream 
is confirmed. The planet-host targets show a kinematic behaviour similar to that of the metal-rich comparison 
subsample, rather than to that of the comparison sample as a whole, thus supporting a primordial origin for the metal 
excess observed in stars with known planetary companions. According to the scenarios proposed as an explanation 
for the dynamical streams, systems with giant planets could have formed more easily in metal-rich inner
 Galactic regions and then been brought into the solar neighbourhood by dynamical streams. 
	  }
\maketitle

\section{Introduction}

Precise spectroscopic studies have revealed that stars with planets seem to be particularly metal-rich 
when compared with ``single'' field dwarfs \citep[e.g.][]{Gonzalez1997,Gonzalezetal2001,Santosetal2001,
Santosetal2004b,Santosetal2005}. This represents the first strong link between the planetary companion 
and its parent star. In the last few years we have published a series of papers on chemical abundances 
in stars with known giant planets. Abundances of volatile and refractory elements in a large set of 
planet-host stars and in a volume-limited comparison sample were presented in \citet{Ecuvillonetal2004a, 
Ecuvillonetal2004b, Ecuvillonetal2006a, Bodagheeetal2003, Beiraoetal2005, Gillietal2006}. The 
volatile/refractory abundance ratios have recently been investigated in \citet{Ecuvillonetal2006b}. Be and Li light 
elements were also analysed \citep{Santosetal2002b,Santosetal2004c, Israelianetal2004}. We are now 
interested in investigating the kinematic behaviour of planet-host stars.

The kinematics of stars in the solar neighbourhood can give fundamental information for our understanding 
of the structure and evolution of the Milky Way. ESA's astrometric satellite {\it Hipparcos} (ESA 1997) 
has provided us with accurate positions and trigonometric parallaxes, as well as absolute proper motions, 
for a large and homogeneous sample of tens of thousands of stars near the Sun. This has offered the 
opportunity to investigate the velocity distribution in the solar neighbourhood, not only for early-type 
stars, but also for the old population of the Galactic disc. Several studies have been performed on this 
topic since \citep[e.g.,][]{Dehnen&Binney1998,Dehnen1998, Alcobe&Cubarsi2005, Chereul&Creze&Bienayme1999, 
Hoogerwerf&Aguilar1999, Asiainetal1999, Mignard2000}. 

The kinematic distribution of stars in the Solar neighbourhood is far from smooth. The inhomogeneities 
can be spatially confined groups of young stars (e.g.\ young clusters), or spatially extended groups with 
the same kinematics. \citet{Eggen1994} defined a ``supercluster'' (SC) as  a group of gravitationally 
unbound stars that share the same kinematics and may occupy extended regions in the Galaxy, 
and a ``moving group'' 
(MG) as the part of the supercluster that enters the solar neighbourhood and can be observed all over the sky. 

According to the standard interpretation, stars in a moving group were formed simultaneously in a small 
phase-space volume, so that we can still observe a stream of young stars with similar velocities. On the 
other hand, old stellar populations should be completely mixed and present a smooth distribution function, 
due to phase mixing and scattering processes. Actually, there are basically two factors acting against the
 persistence of a moving group in the general stellar background: Galactic differential rotation, which 
tends to spread them out very quickly in the direction of the Galactic rotation \citep{Woolley1960}, and 
disc heating \citep[e.g.][]{Lacey1991}, which produces an increase in the velocity dispersion of disc stars 
with age.\citet{Wielen1977} reported that a typical star in the solar neighborhood would forget its initial peculiar velocity within $2\times10^8$ years. 
  
However, several works have obtained results that raise problems with this interpretation. On the one hand, it 
has been reported that early-type stars belonging to the same supercluster span a wide range of ages, in 
contradiction with the assumption of a common origin 
\citep{Chereul&Creze&Bienayme1998,Chereul&Creze&Bienayme1999}.
On the other hand, it is striking to verify that some of the classical MGs are several 10$^8$ yr 
old, and that significant clumpiness in velocity space has been reported for late-type stars \citep{Dehnen1998}.    
Other mechanisms of different nature have been proposed as responsible for the substructures observed in 
velocity space.  
Streams could be the remnants of merger events between 
the Milky Way and a satellite galaxy, as recently proposed by \citet{Navarroetal2004} for the Arcturus 
group, although the chance that two of them have left such important signature in the disc near the Sun 
is statistically very low\citep{Famaeyetal2005a}. Moreover, \citet{Raboudetal1998} reported that the 
local anomaly found in the ($u,v$) plane is caused by intermediate- to high-metallicity stars, while a 
merger would involve rather metal-poor stars \citep{Dehnen2000}. 

Another possibility is that purely dynamical mechanisms, such as non-axysimmetric components of the gravitational potential (for example, the rotating galactic bar or spiral waves), are the cause of substructures in the velocity 
distribution \citep[e.g.][]{Dehnen1998, Raboudetal1998, Sellwood&Binney2002}. Stars in an MG could have 
formed in the inner part of the disc and have become trapped in resonant orbits. 
\citet{Dehnen1999} associates the bimodality of the local velocity distribution of late-type disc stars, lately identified as the 
Hercules stream, with high-metallicity stars thrown out from the inner disc, and \citet{Dehnen2000} concluded that 
outer Lindblad resonant scattering off the Galactic bar is presently the only viable explanation for it. Other simulations have tested the effect of other non-axysimmetric components, in 
particular transient spiral waves, in the Galactic disc \citep[e.g.][]{Sellwood&Binney2002, Desimoneetal2004}, 
with the result that they may cause radial migration near their corotation radius. 

It is not unlikely that known exoplanetary systems could also have suffered such radial displacements. 
The typical contributors to the kinematic structures in the solar neighbourhood are metal-rich old 
stars \citep{Raboudetal1998}. The well-known metal-rich nature of planet-host stars \citep[e.g.][]{Gonzalez1997,Gonzalezetal2001,Santosetal2001, Santosetal2004b,Santosetal2005} might thus be related with dynamical streams. It is probable 
that planet-host stars may have formed in protoplanetary clouds with a high metal content, such as
 those located in inner regions of the Milky Way, and then be brought closer to the Sun by dynamical
 streams \citep{Famaeyetal2005b}. \citet{Famaeyetal2005b} have found an indicative clump of stars hosting 
planets from \citet{Santosetal2003} in the Hyades streams. Several works have investigated the kinematics of stars with extrasolar planets \citep[e.g.][]{Gonzalez1999, Reid2002, Barbieri&Gratton2002, Santosetal2003}, without finding any significative peculiarity.

Two main scenarios have been proposed to explain the metal excess found in stars with exoplanets. The 
``self-enrichment'' hypothesis suggests that the infall of a large amount of protoplanetary rocky 
material on to the parent star may have enhanced its primordial metallicity \citep{Gonzalez1997,Laughlin2000}.
 In contrast, \citet{Santosetal2000,Santosetal2001} have proposed a ``primordial'' origin for the iron 
overabundance of planet-host stars, so that planetary system formation would be triggered by a high metal 
content of the primordial cloud. Several works have obtained results supporting the ``primordial'' scenario 
\citep{Pinsonneaultetal2001,Santosetal2004b,Reid2002, Sadakaneetal2002,Ecuvillonetal2004a,Ecuvillonetal2004b, 
Ecuvillonetal2006a, Ecuvillonetal2006b, Gillietal2006}. However, some clear evidence of pollution events have 
been reported \citep{Gonzalez1998,Laws&Gonzalez2001,Smithetal2001,Israelianetal2001,Israelianetal2003}, 
so that the ``self-enrichment'' scenario cannot be completely discarded \citep{Vauclair2004}.

This article studies the kinematics of planet-host stars and their relation with the dynamical 
streams in the solar neighbourhood and investigates whether it corresponds to the behaviour of the 
metal-rich population or not. The idea is that if the metal-rich nature of planet-host stars is 
intrinsic, then they should be kinematically identical to the metal-rich comparison sample, while
 if the metallicity excess is due to planet-related enrichment, then they should be kinematically 
identical to the total comparison sample. 

We present accurate kinematics for all the stars belonging to the CORALIE planet search survey 
\citep[see][]{Udryetal2000}. Precise radial velocity measurements and CCF parameters were 
extracted from this unique database, whereas HIPPARCOS \citep{Hipparcos1997} parallaxes and photometry, 
and Tycho-2 \citep{Hogetal2000} proper motions were adopted. The location of stars with planets 
in the thin or thick discs is analysed using both kinematic and chemical constraints. We study 
the kinematic behaviour of known planet-host targets with respect to the remaining stars belonging 
to the kinematically unbiased volume-limited sample and we investigate possible links between 
the presence of planets, the stellar metallicity and the dynamical streams Hyades, Sirius and Hercules.

\begin{figure}
\centering 
\includegraphics[width=8cm]{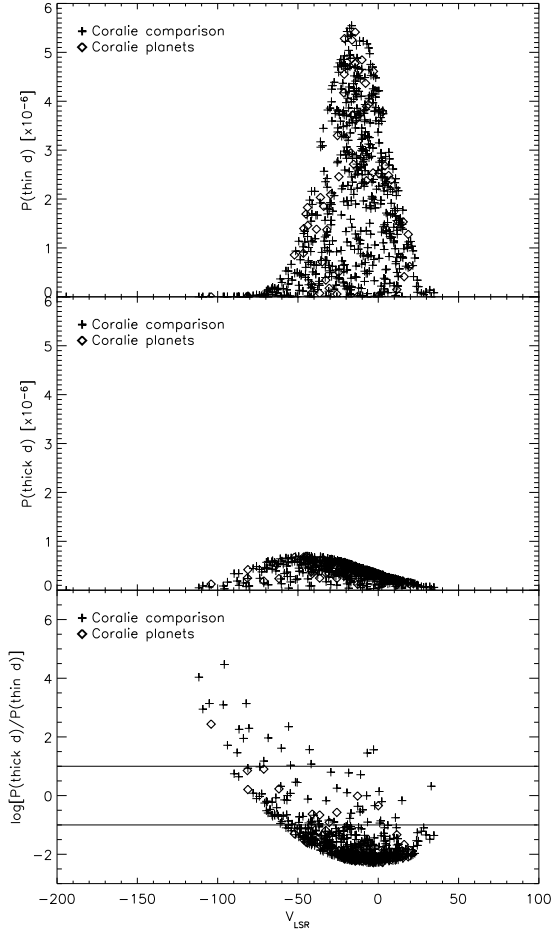}
\caption{Probability distributions of location in the thin disc ({\it upper panel}) and  the 
thick disc ({\it central panel}), and the thick-disc-to-thin-disc relative probability  ({\it 
lower panel}). Diamonds and crosses represent stars with and without planets, respectively. In 
the lower panel, the solid lines indicate the limits for reliable membership to each population.}
\label{ProbDist}
\end{figure}

\begin{figure}
\centering 
\includegraphics[width=8cm]{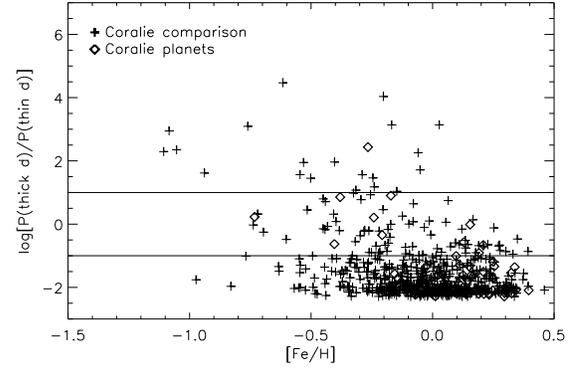}
\caption{Thick-disc-to-thin-disc relative probability  as a function of [Fe/H]. Diamonds and 
crosses represent stars with and without planets, respectively. The solid lines indicate the 
limits for the reliable membership to each population.}
\label{MetProb}
\end{figure}

\section{Data and analysis} 

\begin{figure}
\centering 
\includegraphics[width=6.7cm]{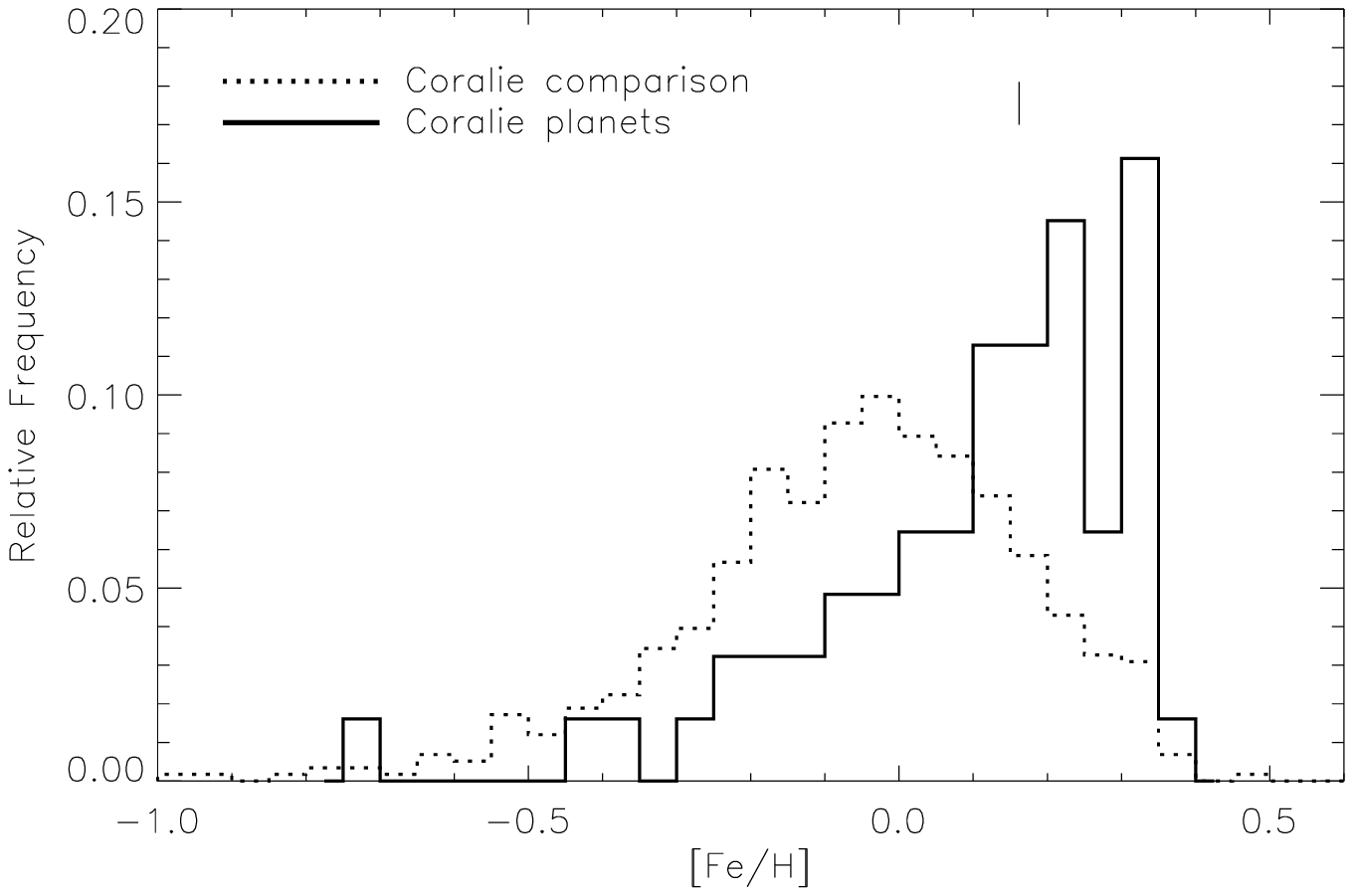}
\includegraphics[width=6.7cm]{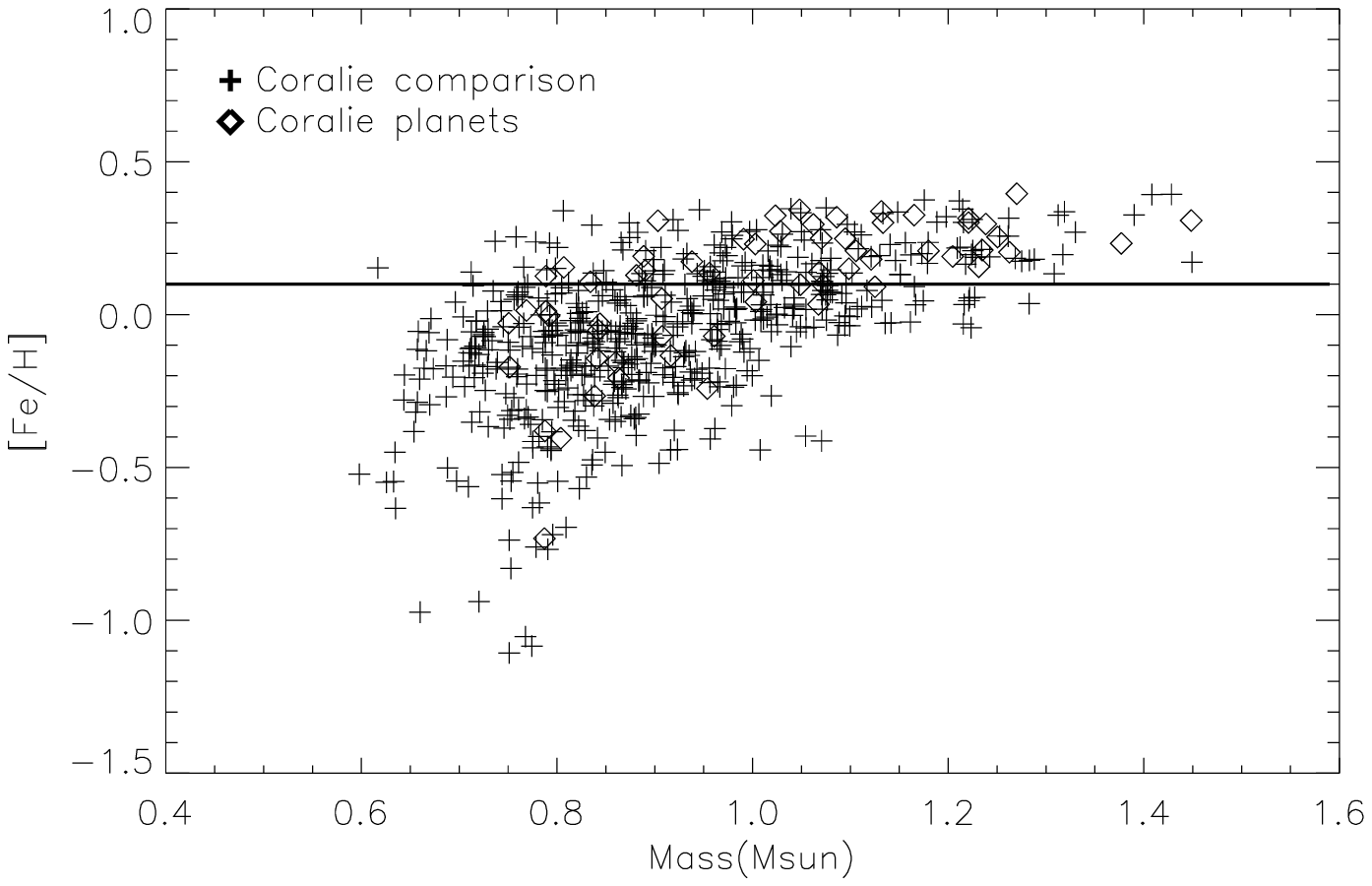}
\caption{{\it Upper panel}: Metallicity distributions of planet-host and comparison stars 
(solid and dotted lines, respectively). The mark indicates the median of the planet-host 
distribution. {\it Lower panel}: [Fe/H] vs.\ stellar mass (in solar mass units) for planet-host 
({\it diamonds}) and comparison ({\it crosses}) stars.}
\label{MetDist}
\end{figure}

\begin{table}[!]
\caption[]{Kinematic parameters adopted for the Hyades, Sirius and Hercules velocity ellipsoids. 
The mean velocities ($U_0$,$V_0$,$W_0$) are relative to the LSR. All values are expressed in km\,s$^{-1}$.}
\begin{center}
\begin{tabular}{c|rrr}
\noalign{\smallskip}
(km\,s$^{-1}$) & Hyades & Sirius & Hercules \\ 
\hline
\noalign{\smallskip}
$U_0$      & $-$20 & 3    & $-$62 \\
$U$ semiaxis & 15    & 10   & 12    \\
$V_0$      & $-$12 & $-$1 & $-$49 \\
$V$ semiaxis & 10    & 7    & 10    \\
$W_0$      & $-$2  & $-$5 & $-$12 \\
$W$ semiaxis & 15    & 13   & 10    \\
\noalign{\smallskip}
\end{tabular}
\end{center}
\label{def1}
\end{table}

\begin{table}[!]
\caption[]{Number of targets falling into the velocity ellipsoids adopted for the Hyades, Sirius and Hercules streams.}
\begin{center}
\begin{tabular}{l|cccc}
\noalign{\smallskip}
 & Field & Hyades & Sirius & Hercules \\ 
\hline
\noalign{\smallskip}
Comparison stars    & 440 & 40 & 16 & 5 \\
Planet-host stars   & 47  & 10 & 1 & 0 \\
\noalign{\smallskip}
\end{tabular}
\end{center}
\label{numdef1}
\end{table}
 
We obtained kinematics for all the stars belonging to the CORALIE planet search survey. 
The overall sample consists of about 1650 unevolved solar-type stars closer than 50 pc, 
with precise parallaxes ($\sigma_{\Pi}$\,$\leq$\,5\,mas) and spectral types between F8 and M1 
in the HIPPARCOS catalogue. A further selection 
was applied to exclude binaries ($\sigma_{RV}$\,$<$\,500\,m\,s$^{-1}$). 
We used parallax measurements and photometric data from the HIPPARCOS catalogue \citep{Hipparcos1997}, 
while coordinates and stellar motions were extracted from Tycho-2 catalogue \citep{Hogetal2000}, 
since they improved the HIPPARCOS data by combining several new catalogues. The CORALIE database 
provided us with radial velocity measurements of a unique precision, as well as with the average 
width and depth of the cross correlation function (CCF).
We derived Galactic space-velocity components $U$, $V$ and $W$ with respect to the Local Standard of 
Rest (LSR), adopting the standard solar motion ($U_{\odot}$,$V_{\odot}$,$W_{\odot}$) = (10.00, 
5.25, 7.17) km\,s$^{-1}$ from \citet{Dehnen&Binney1998}. 

\subsection{Thin/thick disc}
We selected thin and thick disc stars 
from our sample following the procedure described by \citet{Bensbyetal2003}. We assumed that the 
Galactic space velocities of each population (thin disc, thick disc and halo stars) have Gaussian 
distributions. The characteristic velocity dispersions ($\sigma_U$, $\sigma_V$, $\sigma_W$) and 
asymmetric drift ($V_{\rm asym}$) were taken from \citet{Bensbyetal2003}, as well as the observed 
fractions of each population in the solar neighbourhood ($X$). The probability distribution of 
belonging to a specific population ($P$) is thus given by the Gaussian probability distribution 
multiplied by the corresponding observed fraction ($X$). Stars with $P$(thick disc)/$P$(thin 
disc)\,$\geq10$ and $\leq0.1$ will be reliable members of the thick and thin disc, respectively. 
Figure~\ref{ProbDist} shows the probability distributions of belonging to the thin disc ({\it upper 
panel}) and to the thick disc ({\it central panel}), and the thick-disc-to-thin-disc relative probability  
({\it lower panel}). 
Chemical abundances for CORALIE stars already spectroscopically studied were extracted from 
\citet{Ecuvillonetal2004a, Ecuvillonetal2004b, Ecuvillonetal2006a} and \citet{Gillietal2006} and 
combined with the kinematic information in order to analyse the relation of the trends [$X$/Fe] vs.\ [Fe/H] 
with the thin and thick discs.  

All the metallicities were derived using the calibration of the CCF surface with [Fe/H] by 
\citet{Santosetal2002a}. The distribution of the thick-disc-to-thin-disc relative probability  
as a function of metallicity is shown in Figure~\ref{MetProb}. We also computed stellar masses 
by a comparison of the star positions in the CM diagram with the calculated isochrones of 
\citet{Pietrinfernietal2004}. Ages for field stars are notoriously difficult to determine, 
and suffer from large biases that make them unsuitable for this kind of study, as discussed by 
\citet{Pont&Eyer2004}. However, mass can be used as an age surrogate for a first-order 
separation of young and old stars. Masses can be determined quite precisely from stellar evolution 
models and are less subject to biases. Figure~\ref{MetDist} shows the metallicity distributions for 
planet-host and comparison stars ({\it upper panel}), and the plot of [Fe/H] vs.\ mass ({\it lower panel}).

\begin{figure}
\centering 
\includegraphics[width=8cm]{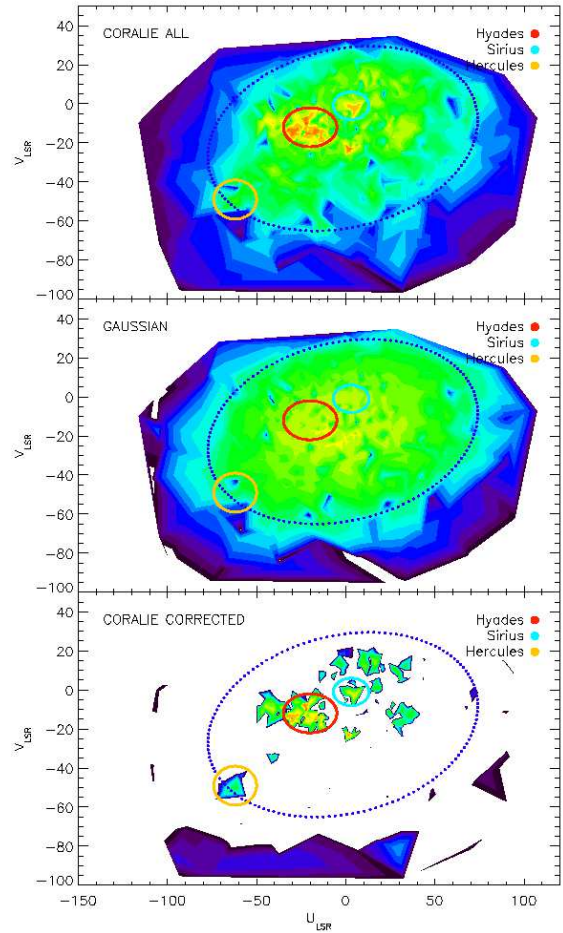}
\caption{Density distribution of Galactic space velocities projected on the ($V,U$) plane, for 
the whole comparison sample. The uncorrected density distribution, the subtracted Gaussian distribution 
and the corrected density distribution are presented in the upper, central and lower panels, 
respectively. The field ellipsoid is overplotted with the blue line, while the Hyades, Sirius and 
Hercules ellipsoids are indicated by the red, blue green and yellow lines.}
\label{VvsUMTot}
\end{figure}

\begin{figure}
\centering 
\includegraphics[width=8cm]{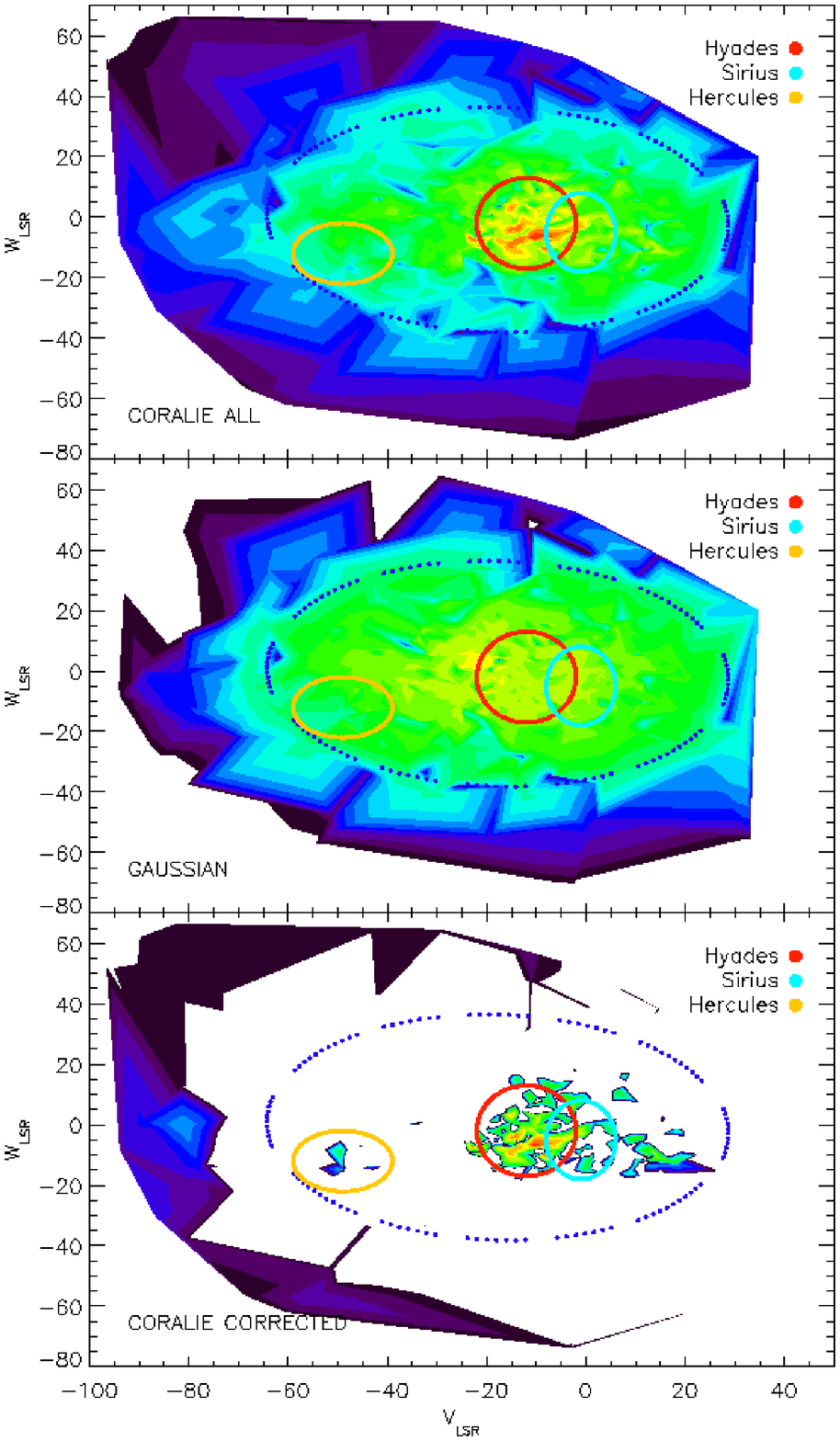}
\caption{Density distribution of Galactic space velocities projected on the ($W,V$) plane for the
 whole comparison sample. The uncorrected density distribution, the subtracted Gaussian distribution 
and the corrected density distribution are presented in the upper, central and lower panels, 
respectively. The field ellipsoid is overplotted with the blue line, while the Hyades, Sirius and Hercules 
ellipsoids are indicated by the red, blue green and yellow lines.}
\label{WvsVMTot}
\end{figure}

\begin{figure}
\centering 
\includegraphics[width=8cm]{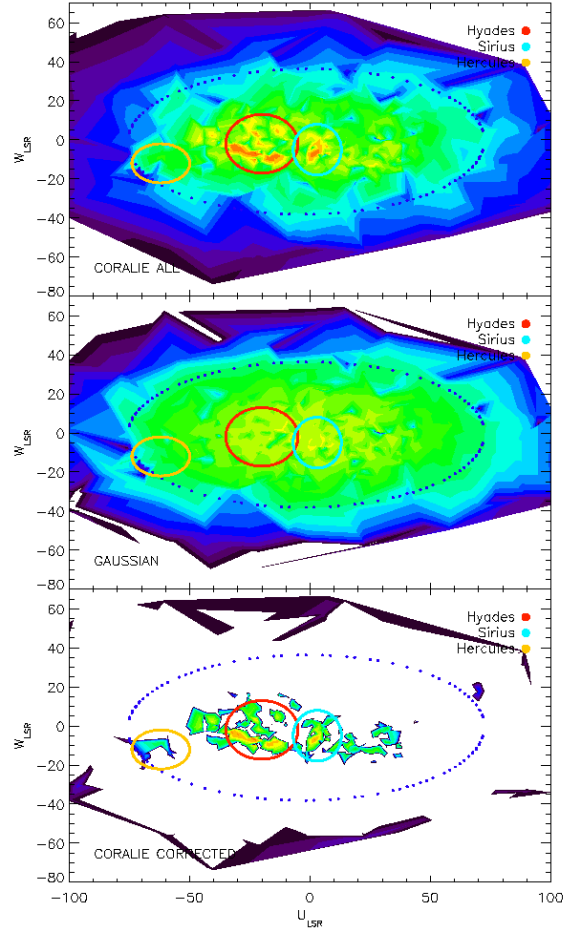}
\caption{Density distribution of Galactic space velocities projected on the ($W,U$) plane for the whole 
comparison sample. The uncorrected density distribution, the subtracted Gaussian distribution and the 
corrected density distribution are presented in the upper, central and lower panels, respectively. The 
field ellipsoid is overplotted with the blue line, while the Hyades, Sirius and Hercules ellipsoids are 
indicated by the red, blue green and yellow lines.}
\label{WvsUMTot}
\end{figure}

\subsection{Analysis of the ($U, V, W$) space}
The field ellipsoid was defined by computing its inertial tensor: the eigenvectors of the inertial tensor would indicate the main directions 
($U', V', W'$) of the field ellipsoid in the ($U, V, W$) space. We adopted 2\,$\sigma$ of the data along the 
directions $U'$, $V'$ and $W'$ as the field ellipsoid semiaxes. We then derived the density distribution of 
the Galactic space velocities by the fifth nearest-neighbour method. This method assigns to each point a 
density value, according to its surroundings \citep[see][section 5.2]{Silverman1996}. 

A field without ``structures'' is supposed to follow a Gaussian distribution. Therefore, this Gaussian 
``background'' has to be removed to study the remnant ``structures''. We thus subtracted from the field 
density distribution the density distribution of the Gaussian corresponding to the previously defined 
field ellipsoid. Figures~\ref{VvsUMTot},~\ref{WvsVMTot} and~\ref{WvsUMTot} illustrate the density 
distribution of Galactic space velocities projected on the ($V,U$), ($W,V$) and ($W,U$) planes, respectively. 
The uncorrected distributions are shown in the upper panels, while the subtracted Gaussian distributions 
and the corrected density distributions are presented in the lower panels. 

The next step consisted in defining the ellipsoids describing the remnant ``structures'' in the corrected 
density distribution. As a first aproximation, we adopted the kinematic parameters of the Hyades, Sirius and 
Hercules streams derived by \citet{Famaeyetal2005a}, and then adjusted by eye the ellipsoid centroids and 
semiaxes according to what we observed. Table~\ref{def1} lists the kinematic parameters finally adopted for
 each velocity ellipsoid. In Figures~\ref{VvsUMTot},~\ref{WvsVMTot} and~\ref{WvsUMTot} the projected 
ellipsoids are represented and identified as the Hyades, Sirius and Hercules streams. The projected field 
ellipsoid is also overplotted in each figure ({\it blue line}). The number of targets falling into each velocity 
ellipsoid are listed in Table~\ref{numdef1}. 

The method used to obtain the field ellipsoid does not make any {\it a priori} assumptions concerning the dynamical streams. A different approach would be to adopt a negligible vertex deviation for the field ellispoid by considering that the Hercules stream is  its main contributor \citep[e.g.][]{Muhlbauer&Dehnen2003}. We have checked that the assumption of a negligible vertex deviation does not change our results. Although the Hercules ellipsoid results slightly enlarged, and consequently the number of comparison stars belonging to the stream is enhanced from 5 to 8, no planet-host stars are found within the modified Hercules ellipsoid.  

Next, we defined the groups in which our sample would be divided for the comparison: planet hosts, 
comparison sample, metal-rich comparison stars and comparison stars with solar metallicity or lower. 
The metallicity cutoff was chosen so that the median of the metal-rich comparison sample was equal 
to the one of the planet-host distribution ([Fe/H]=0.15, see Figure~\ref{MetDist}, {\it upper panel}), and 
also so that the mass-metallicity, or age-metallicity, relation affected as few targets as possible 
(see Figure~\ref{MetDist}, {\it lower panel}). Cutting the comparison sample at [Fe/H] = 0.07 fulfilled 
the condition concerning medians. The metallicity cutoff was then rounded off to [Fe/H] = 0.1.

We then proceeded to count how many stars from each group fell into the ellipsoids corresponding to the 
field and to the Hyades, Sirius and Hercules streams. We also counted the numbers expected in the streams 
if the group were to have followed a Gaussian distribution, with $\sigma$ equal to the velocity dispersion of 
the group. For the planet-host sample, we did this  for two Gaussian distributions, the first with the dispersion 
characteristic of the whole comparison sample, and the second with the dispersion characteristic of the 
metal-rich comparison subsample. The count for each stream was corrected by subtracting the expected 
number from the Gaussian distribution and normalized by dividing by the total number of the group in the 
field. This corresponds to the ratio of excess objects in the stream. The Sirius and Hercules 
streams were excluded from this analysis since the two velocity ellipsoids contained respectively one 
and no planet host stars. 

A further analysis was carried out by means of the density distributions of the Galactic space 
velocities, derived by using the fifth nearest-neighbour method. For each group independently, we computed 
the density distribution corresponding to a Gaussian distribution with $\sigma$ equal to the 
velocity dispersion of the group. In the case of the planet-host group, two Gaussian distributions 
were tested, one with the dispersion of the whole comparison sample, and the other with the dispersion 
of the metal-rich comparison subsample. The ``observed'' distribution was then corrected by the Gaussian 
contribution. 
From the corrected density distribution, we derived an average density for the field and the Hyades 
ellipsoids. The averaged field density was then subtracted that of the Hyades in order to obtain the 
overdensity over the mean field ``level''. 

\begin{figure}
\centering 
\includegraphics[width=7cm]{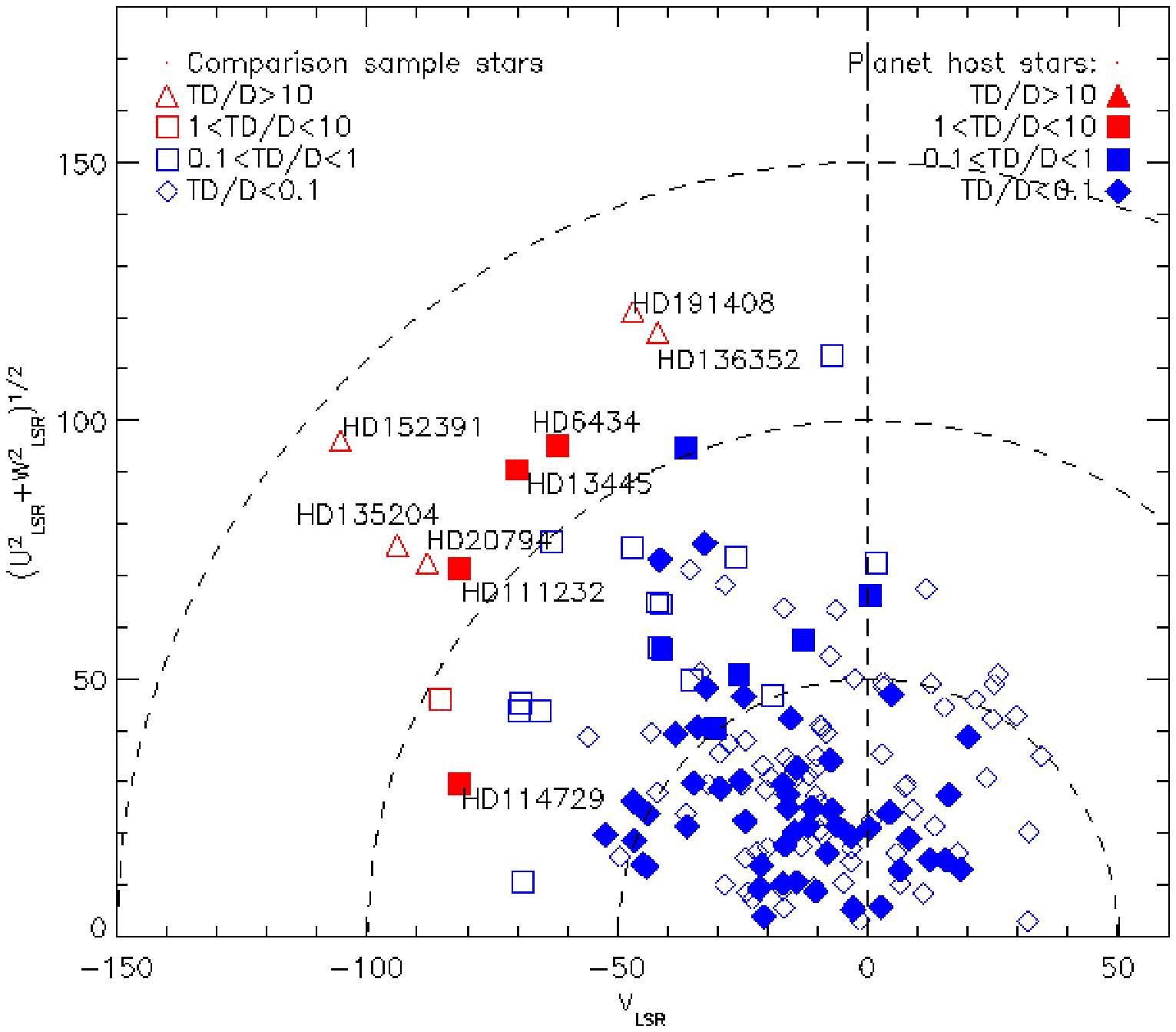}
\caption{Toomre diagram for the CORALIE targets with chemical abundances by \citet{Ecuvillonetal2004a, 
Ecuvillonetal2004b, Ecuvillonetal2006a} and \citet {Gillietal2006}. Dotted lines indicate constant
 peculiar velocities $v_{\rm pec}=(U^2_{\rm LSR}+V^2_{\rm LSR}+W^2_{\rm LSR})^{1/2}$ in steps of 50\,km\,s$^{-1}$. 
Filled and open symbols represent stars with and without planets, respectively. Targets that are full 
and intermediate members of the thick disc population are indicated with red triangles and squares, 
respectively, while full and intermediate members of the thin disc are represented by blue diamonds 
and squares, respectively. The four comparison targets full members, and the four planet-host stars 
intermediate members of the thick disc, are identified by their numbers.}
\label{toomre}
\end{figure}

\begin{figure*}
\centering 
\includegraphics[width=7cm]{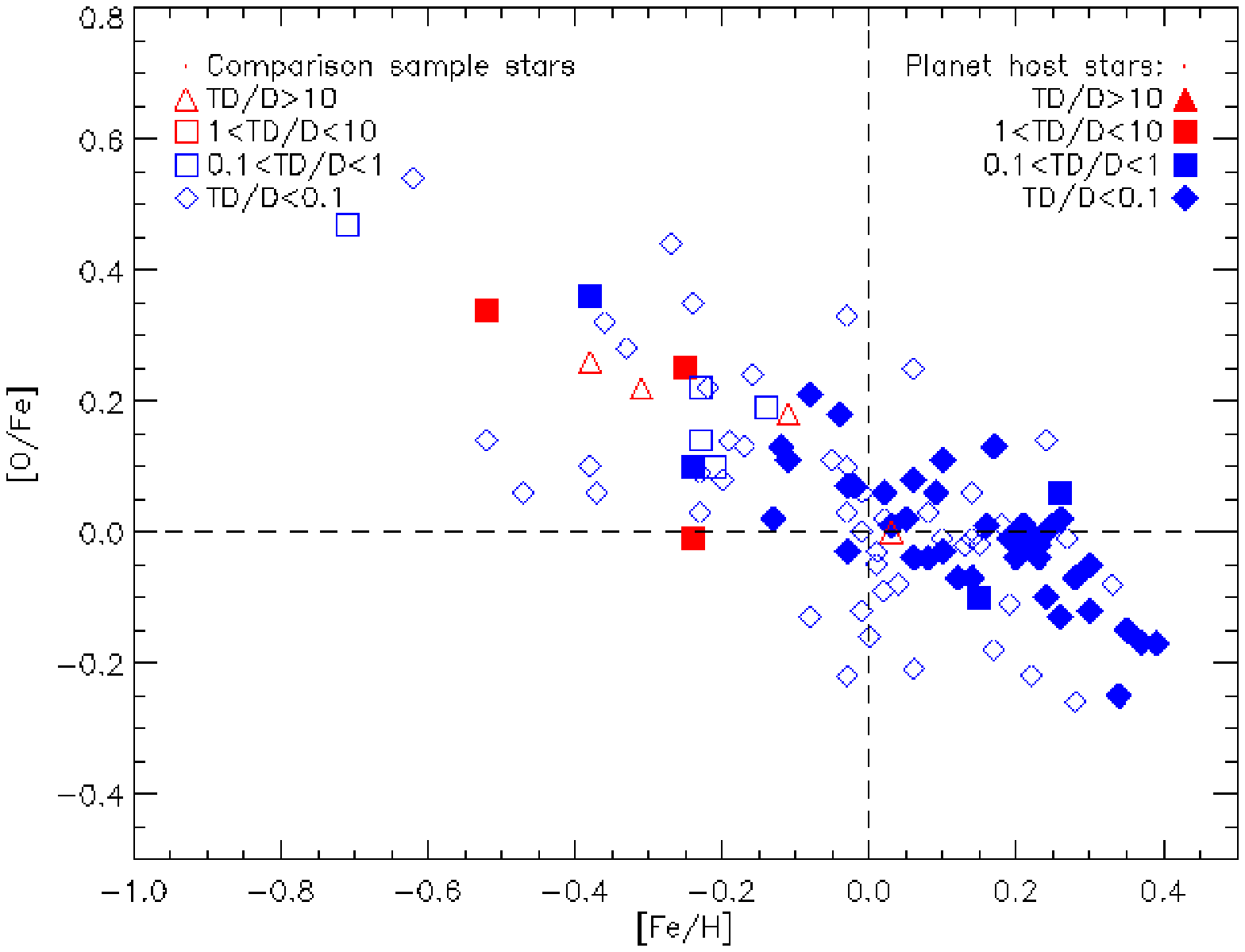}
\includegraphics[width=7cm]{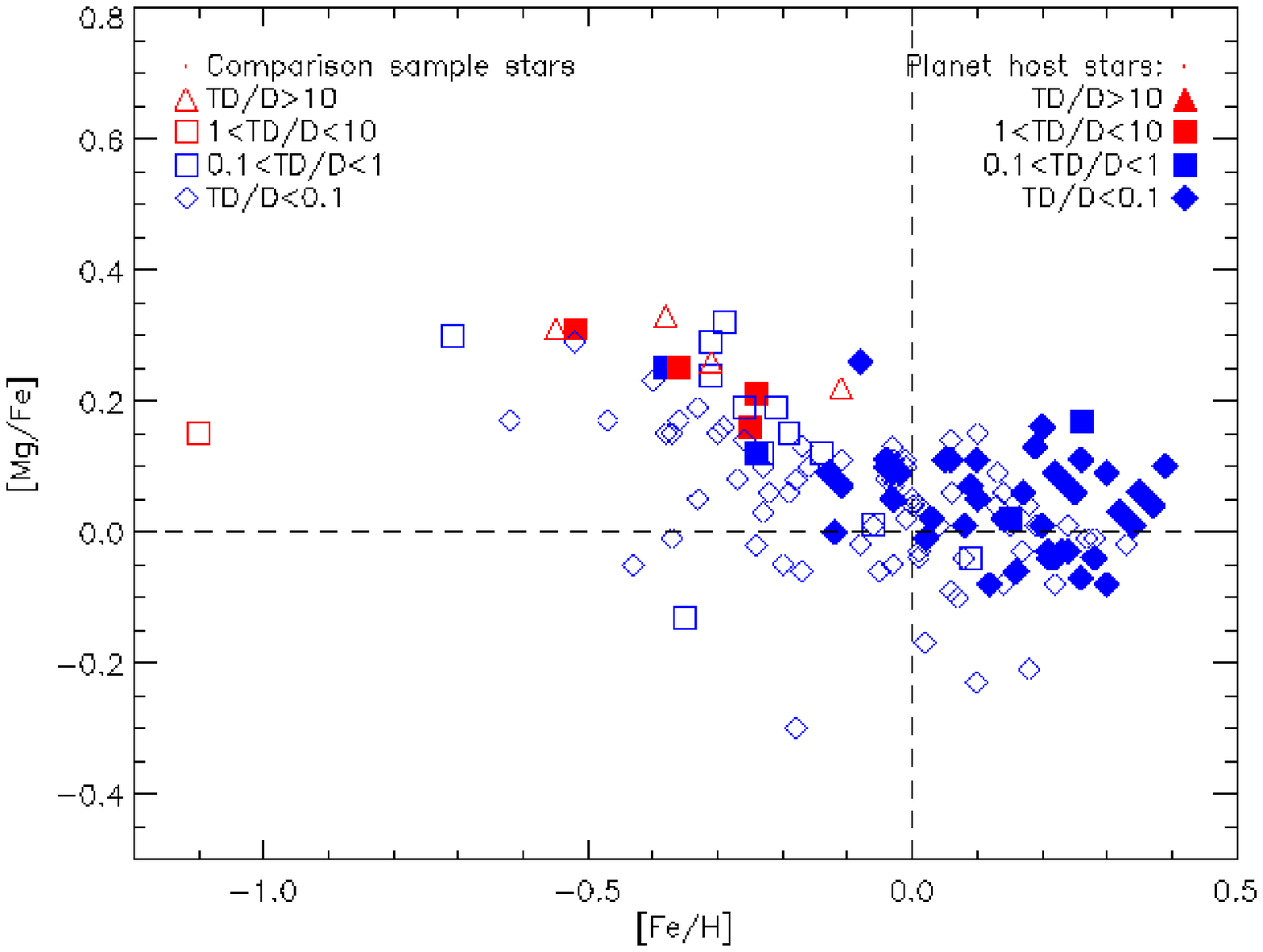}
\caption{[O/Fe] and [Mg/Fe] vs.\ [Fe/H] for CORALIE targets with chemical abundances by 
\citet{Ecuvillonetal2004a, Ecuvillonetal2004b, Ecuvillonetal2006a} and \citet {Gillietal2006}.
 Filled and open symbols represent stars with and without planets, respectively. Targets which
 are full and intermediate members of the thick disc population are indicated with red triangles 
and squares, respectively, while full and intermediate members of the thin disc are represented 
by blue diamonds and squares, respectively.}
\label{OxyMg}
\end{figure*}

\begin{figure*}
\centering 
\includegraphics[width=14cm]{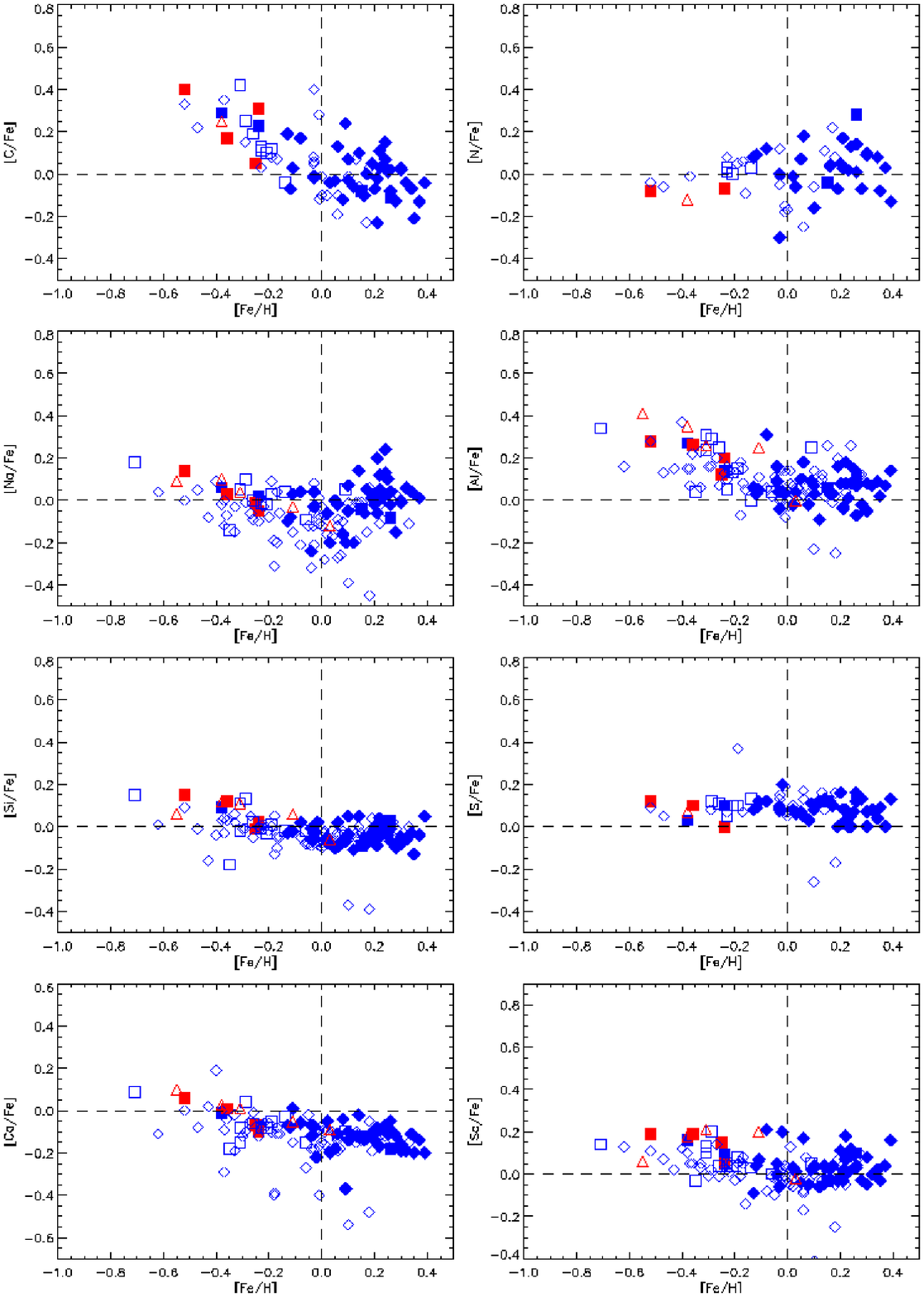}
\caption{[$X$/Fe] vs.\ [Fe/H] (from C to Sc) for CORALIE targets with chemical abundances by 
\citet{Ecuvillonetal2004a, Ecuvillonetal2004b, Ecuvillonetal2006a} and \citet {Gillietal2006}. 
Filled and open symbols represent stars with and without planets, respectively. Targets which 
are full and intermediate members of the thick disc population are indicated with red triangles 
and squares, respectively, while full and intermediate members of the thin disc are represented by 
blue diamonds and squares, respectively.}
\label{multi}
\end{figure*}

\begin{figure*}
\centering 
\includegraphics[width=14cm]{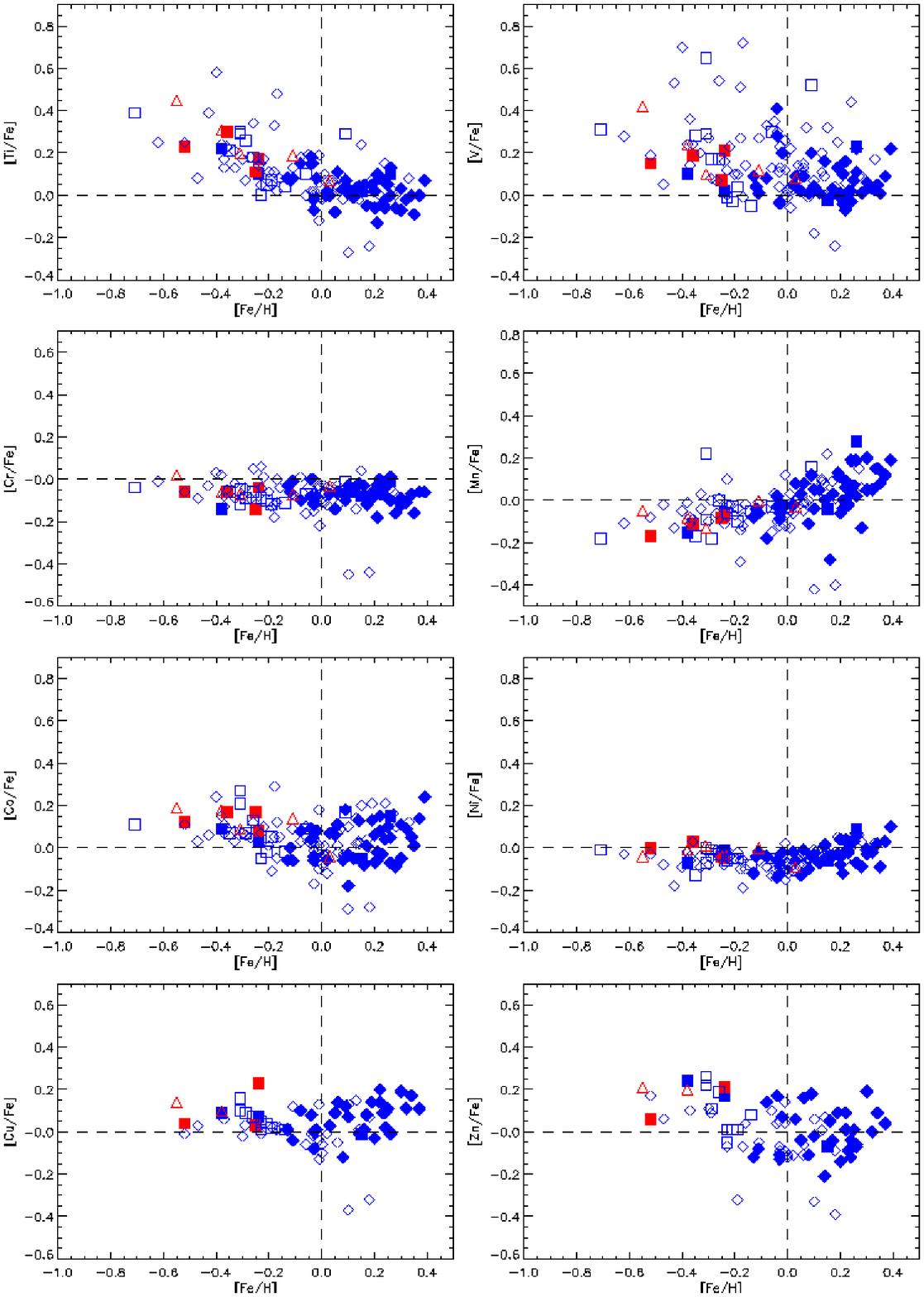}
\caption{Continuation of Figure~\ref{multi}. From Ti to Zn.}
\label{multicont}
\end{figure*}

\section{Results and discussion}

\subsection{Thin/thick disc}

Our results reveal that the CORALIE sample bears one planet-host target, \object{HD\,4308}, 
and 21 stars with no known planetary companions, with kinematics typical of the thick disc 
(see Figure~\ref{ProbDist}). The metallicities of \object{HD\,4308} ([Fe/H] = $-0.27$ from the 
CCF calibration) and of the thick disc targets without known planets are typical of the thick 
disc population (see Figure~\ref{MetProb}). Only one comparison star, \object{HD\,152391}, 
shows a supersolar metallicity ([Fe/H] = 0.03 from the CCF calibration).
The following study combining both kinematical and chemical information will be limited to 
the subset of CORALIE stars with available homogeneous abundances from \citet{Ecuvillonetal2004a, 
Ecuvillonetal2004b, Ecuvillonetal2006a} and \citet{Gillietal2006}. 

The Toomre diagram for the subset of CORALIE stars with chemical abundances from 
\citet{Ecuvillonetal2004a, Ecuvillonetal2004b, Ecuvillonetal2006a} and \citet{Gillietal2006} 
is shown in Figure~\ref{toomre}. Five comparison stars are very likely members of the thick disc 
population (\object{HD\,20794}, \object{HD\,135204}, \object{HD\,136352}, \object{HD\,152391}, 
\object{HD\,191408}), while no planet-host targets present such a high probability. Only the four 
planet hosts \object{HD\,6434}, \object{HD\,13445}, \object{HD\,111232} and \object{HD\,114729}, 
are potential members of the thick disc, with probabilities $P$(thick disc)/$P$(thin disc) between 1 and 10. 
The [$X$/Fe] vs.\ [Fe/H] trends for oxygen and magnesium are shown in Figure~\ref{OxyMg}, and in 
Figures~\ref{multi} and~\ref{multicont} for the rest of the elements (C, N, Na, Al, Si, S, Ca, Sc, 
Ti, V, Cr, Mn, Co, Ni, Cu and Zn).
  
Previous works \citep[e.g.][]{Fuhrmann1998, Feltzingetal2003, Bensbyetal2003} reported different 
abundance trends for the thin and thick disc populations. In particular, the $\alpha$-elements 
oxygen and magnesium show clearly distinct trends for the thin and thick discs at sub-solar 
metallicities \citep{Bensbyetal2003,Bensbyetal2004}. However, \citet{Chenetal2000} presented 
results from a sample of 90 F and G dwarfs which show no significant scatter in $\alpha$-element 
ratios as function of [Fe/H]. Another interesting issue is the existence of stars with thick disc 
kinematics at high metallicities,  pointed out by several authors \citep[e.g.][]{Bensbyetal2003, 
Misheninaetal2004}.

Our results confirm the possibility of thick disc stars at [Fe/H] $>$ 0, since one of the five 
comparison stars with thick disc kinematics that was previously chemically analysed, \object{HD\,152391}, 
has [Fe/H] = 0.03. The same target has been already reported as thick disc member with supersolar metallicity 
in the kinematic study of the whole CORALIE sample (see the first paragraph of this section). It is 
interesting to note how the metallicity derived by the CCF calibration agrees perfectly with the iron 
abundance issued by the detailed spectroscopic study by \citet{Santosetal2005}. 

Nevertheless, we did not find any difference in the abundance patterns marked by thick and thin 
disc stars, nor any significant signature of enrichment in $\alpha$-elements for thick disc 
population. For magnesium (see Figure~\ref{OxyMg}, {\it right panel}), it might be possible that 
the thick disc members lay on the upper envelope of the points. However, the small number of possible 
members for this population in our sample does not enable us to reach any conclusion. 

\subsection{Dynamical streams}
\label{sub3-2}

\begin{table*}[!]
\caption[]{Results for the ``counting'' method in the Hyades stream. The {\it Definition 1} and 
{\it Definition 2} columns refer to the adopted definitions for the stream ellipsoids, from our work and 
from \citet{Famaeyetal2005a}, respectively. The numbers of stars lying in the field and Hyades 
ellispoids are listed, as well as the expected number of stars for a Gaussian distribution, and the
 resulting ratio of objects in excess in the Hyades stream. The assumed uncertainties are Poissonian errors.}
\begin{center}
\begin{tabular}{c|cccc|cccc}
\noalign{\smallskip}
 & \multicolumn{4}{c|}{Definition 1} & \multicolumn{4}{c}{Definition 2} \\ 
 Group description & Field & Hyades & Exp Hyades & Excess & Field & Hyades & Exp Hyades & Excess \\
 \hline
 \noalign{\smallskip}
 Planet-hosts: $\sigma$ whole comp. 	    & 47  & 10 & 1.7  & 0.18$\pm$0.07 & 47 & 7  & 1.4 & 0.12$\pm$0.05 \\ 
 \multicolumn{1}{r|}{$\sigma$ M-rich comp.} & ''  & '' & 2.3  & 0.16$\pm$0.06 & '' & '' & 1.9 & 0.11$\pm$0.06 \\ 
 Whole comparison                           & 440 & 40 & 16.3 & 0.05$\pm$0.01 & 440 & 33 & 13.0 & 0.05$\pm$0.01 \\
 M-Rich comparison                          & 119 & 19 & 5.7  & 0.11$\pm$0.04 & 119 & 17 & 4.7 & 0.10$\pm$0.03 \\
 Solar comparison                           & 321 & 21 & 11.9 & 0.03$\pm$0.02 & 321 & 16 & 9.2 & 0.02$\pm$0.01 \\
\noalign{\smallskip}
\end{tabular}
\end{center}
\label{test1}
\end{table*}

\begin{table*}[!]
\caption[]{Results for the ``overdensity'' method in the Hyades stream. The {\it Def.1} and {\it Def.2} 
columns refer to the adopted definitions for the velocity ellipsoids, from our work and from 
\citet{Famaeyetal2005a}, respectively. {\it Def.1 mod.1} corresponds to the case without 
substracting the average density of the field, while {\it Def.1 mod.2} reports the results for 
a metallicity cutoff [Fe/H]$_{cutoff}$=0.05 (see the Subsection~\ref{sub3-2} for more details). 
The last column, {\it Def.1 M$<$1.2M$_{\odot}$}, refers to the results obtained by excluding the 
targets with  M$>$1.2M$_{\odot}$. The uncertainties were derived applying the bootstrap method.} 
\begin{center}
\begin{tabular}{c|c|c|c|c|c}
\noalign{\smallskip}
\multicolumn{1}{c}{} & \multicolumn{5}{|c}{Hyades Overdensity} \\
Group description & Def.\ 1 & Def.\ 2 & Def.\ 1 mod.\ 1 & Def.\ 1 mod.\ 2 & Def.\ 1 $M<1.2M_{\odot}$ \\ 
 \hline
 \noalign{\smallskip}
 Planet-hosts: $\sigma$ whole comp. 	    & 0.031$\pm$0.017 & 0.030$\pm$0.017 & 0.031 & 0.031$\pm$0.018 & 0.011$\pm$0.020 \\ 
 \multicolumn{1}{r|}{$\sigma$ M-rich comp.} & 0.029$\pm$0.018 & 0.027$\pm$0.018 & 0.028 & 0.028$\pm$0.017 & 0.009$\pm$0.020 \\ 
 Whole comparison			    & 0.004$\pm$0.004 & 0.004$\pm$0.004 & 0.004 & 0.004$\pm$0.004 & 0.003$\pm$0.004 \\ 
 M-Rich comparison                          & 0.020$\pm$0.011 & 0.019$\pm$0.012 & 0.020 & 0.012$\pm$0.007 & 0.013$\pm$0.011 \\ 
 Solar comparison                           & 0.003$\pm$0.005 & 0.003$\pm$0.004 & 0.003 & 0.002$\pm$0.003 & 0.003$\pm$0.005 \\ 
\noalign{\smallskip}
\end{tabular}
\end{center}
\label{test2}
\end{table*}

\begin{figure*}
\centering 
\includegraphics[width=8cm,height=8cm]{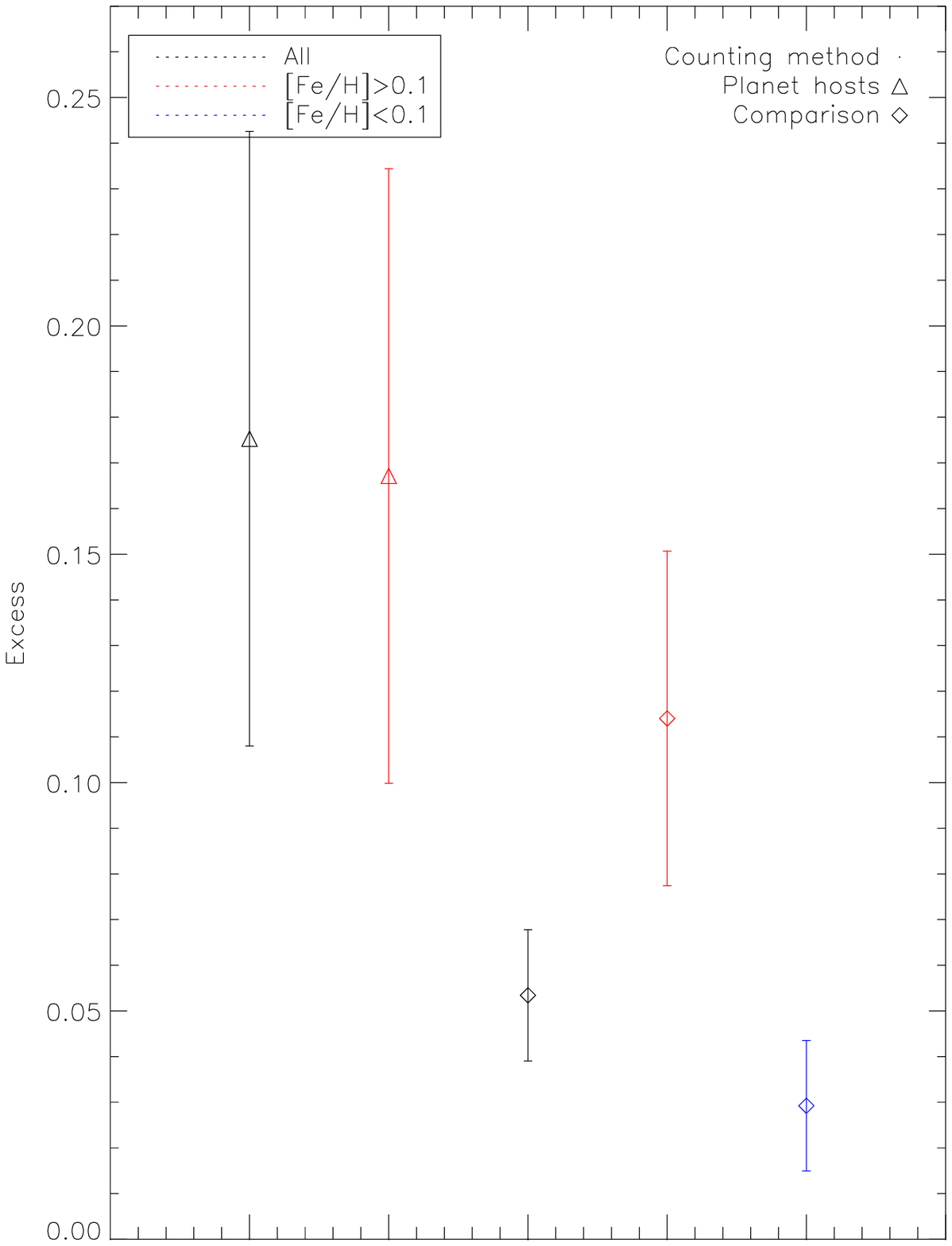}
\includegraphics[width=8cm,height=8cm]{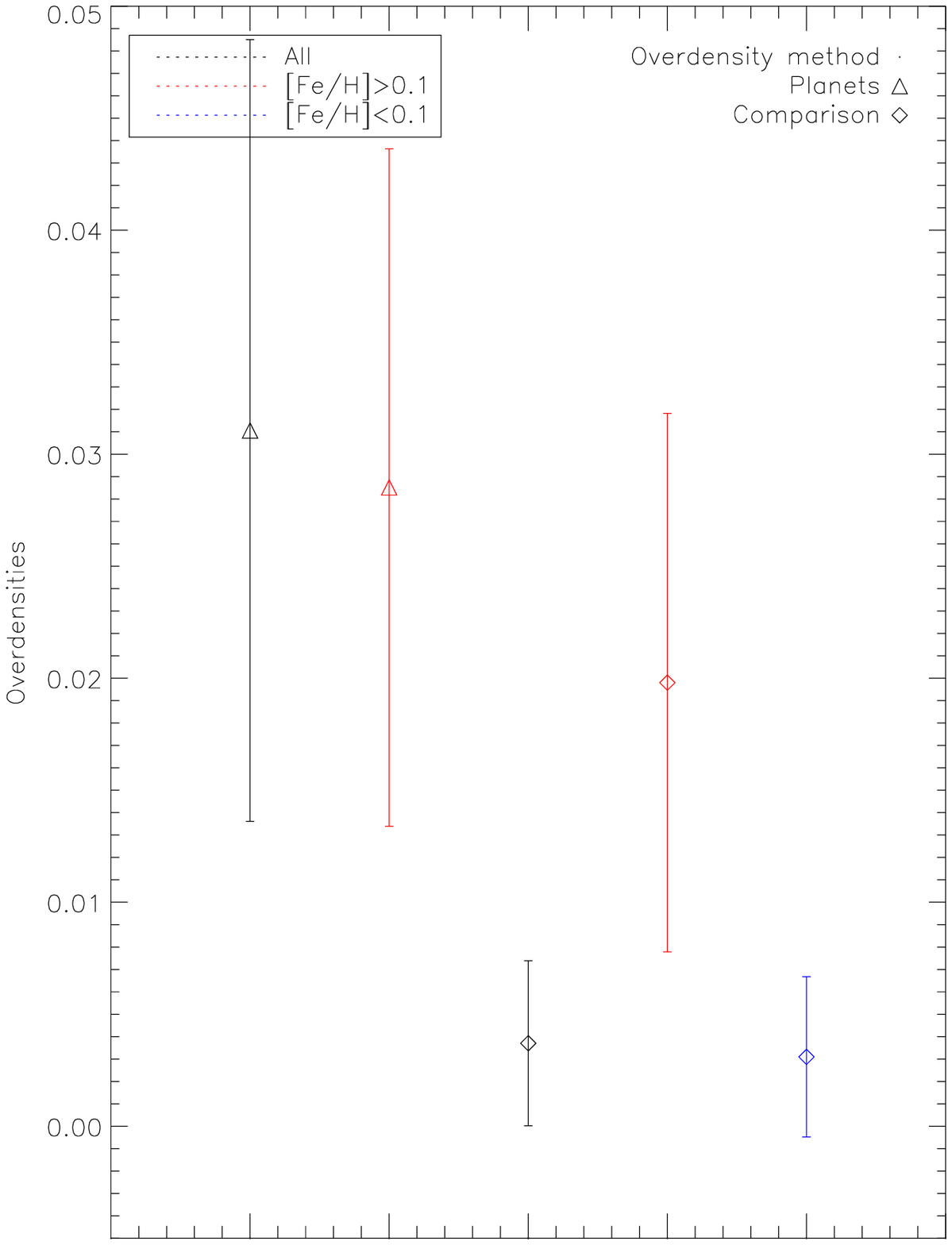}
\caption{Results from the ``counting'' ({\it left panel}) and the ``overdensity'' methods 
({\it right panel}). Black, red and blue diamonds represent the whole comparison sample, 
and the metal-rich and solar subsamples, respectively. Black and red triangles indicate planet-host 
stars, adopting a Gaussian distribution with $\sigma$ equal to the velocity dispersion of the whole 
comparison sample, and of the metal-rich subsample, respectively. The error bars show Poissonian 
errors for the ``counting'' method ({\it left panel}), and uncertainties derived by the bootstrap 
method for the ``overdensity'' analysis ({\it right panel}).}
\label{res}
\end{figure*}

\begin{figure*}
\centering 
\includegraphics[width=8cm]{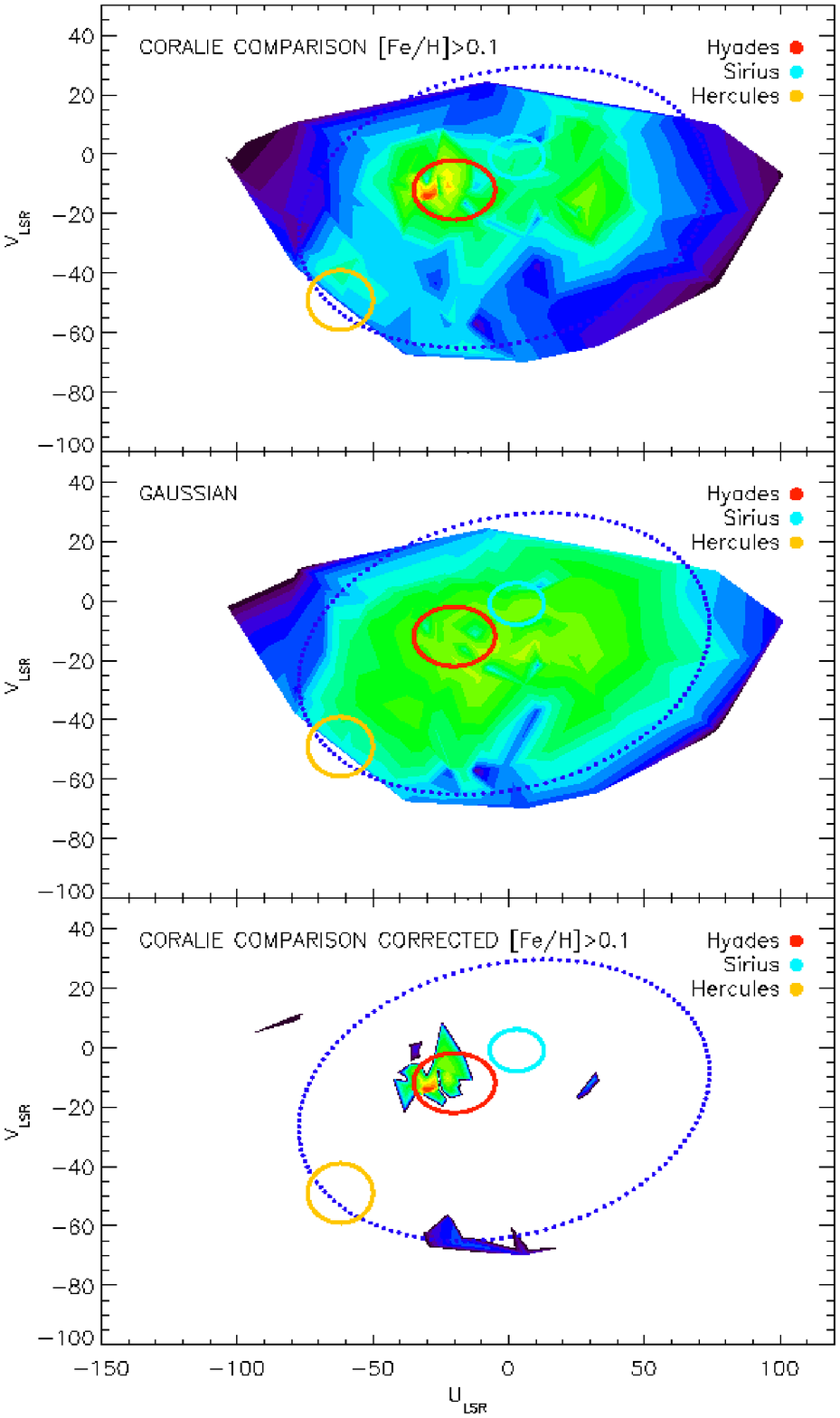}
\includegraphics[width=8cm]{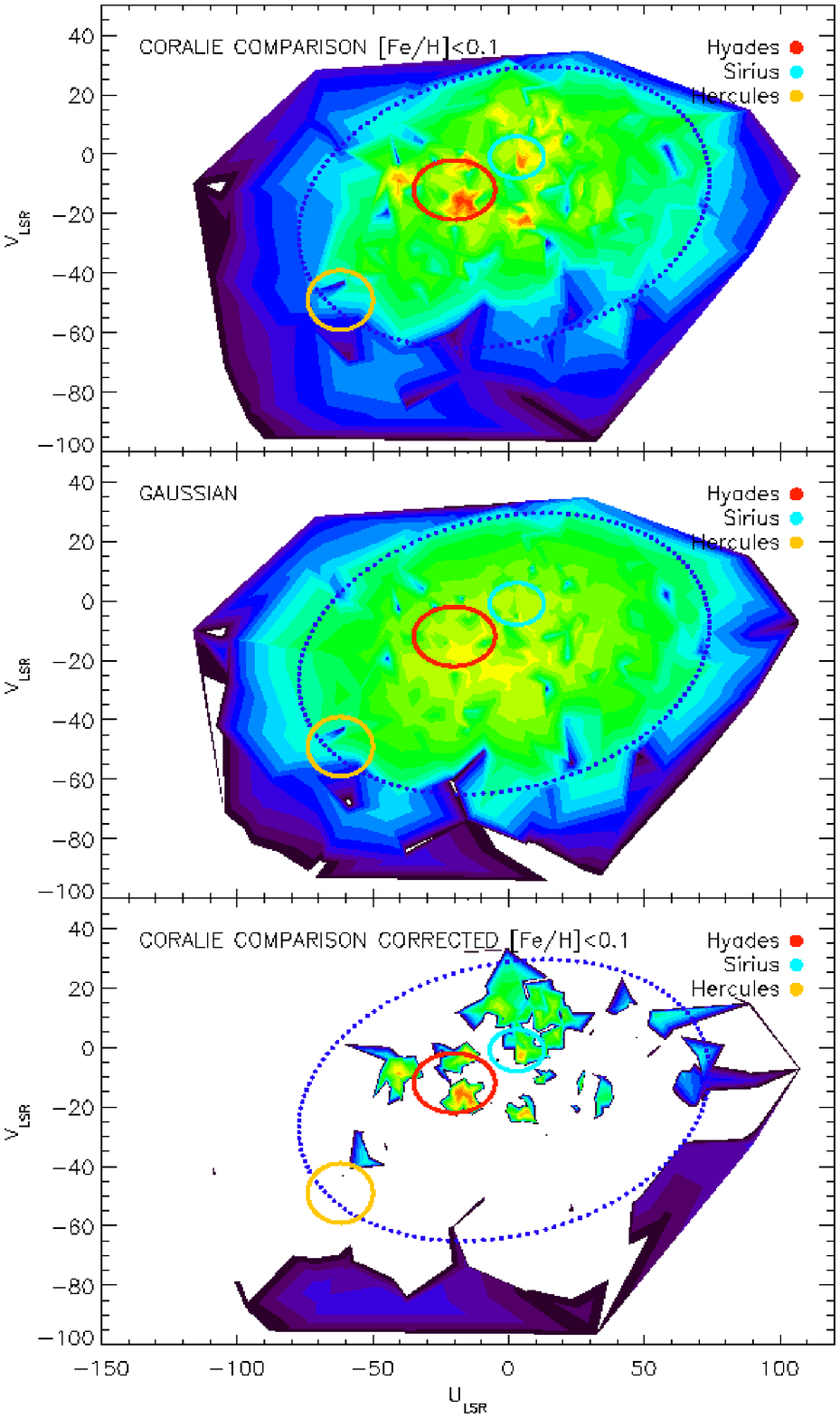}
\caption{Density distributions of Galactic space velocities projected on the ($V,U$) plane, 
for the metal-rich ({\it left panel}) and solar ({\it right panel}) comparison subsamples. 
The uncorrected density distribution, the substracted Gaussian distribution and the corrected 
density distribution are presented in the upper, central and lower panels, respectively. The
 field ellipsoid is overplotted with the blue line, while the Hyades, Sirius and Hercules 
ellipsoids are indicated by the red, blue green and yellow lines.}
\label{MetalPick}
\end{figure*}

Table~\ref{test1} shows the results obtained with the ``counting'' method, using the ellipsoid 
definitions from our work ({\it Definition 1}), and the ones from \citet{Famaeyetal2005b} 
({\it Definition 2}). The first column lists the different groups: planet-host stars, 
assuming for the Gaussian the $\sigma$ of the whole comparison sample, and of the metal-rich 
subsample; the whole comparison sample; and the metal-rich and the solar-metallicity comparison 
subsamples. Then are listed the numbers of targets in the field and Hyades ellipsoids, the 
expected number of targets from the Gaussian distribution, and finally the ratio of objects 
in excess from what expected for a Gaussian distribution. The uncertainties were estimated 
as Poissonian errors.

The results issued from this method indicate that the clumpiness typical of the Hyades is
 much more significant for the metal-rich subsample than for the whole sample. This confirms 
the high average metallicity reported for Hyades stream members by previous studies 
\citep[e.g.][]{Raboudetal1998, Chereul&Grenon2001} and agrees with the mechanism proposed 
as an explanation of the anomalies in the ($U,V$) plane \citep{Dehnen2000, Famaeyetal2005a}. 
According to this description, the stars would have formed at inner Galactic radii, with 
higher metallicity, and then be trapped into resonant orbits by the non-axysimmetric potential, 
and thrown closer to the Sun.

The planet-host group shows an excess in the Hyades stream much more significant than the 
other groups. Errors associated with stars with planets are large because of the small number of 
targets with planetary companions involved in the whole volume-limited sample CORALIE. 
However, the targets with planets present an excess in the Hyades stream significantly 
larger than that obtained for the other groups. In particular, the clumpiness of planet-host 
stars is different from that of the whole comparison sample, with a significance of more than 1.5\,$\sigma$.

The kinematic behaviour of CORALIE stars with planets seems to match better with that 
observed in the metal-rich comparison subsample than with the whole comparison sample. Both
 metal-rich and planet-host samples show a much larger clumpiness  than the whole comparison sample. Nevertheless, it is important to note that the presence of planet-host stars in the 
Hyades stream is even denser than that found for metal-rich ``single'' stars. This outcome 
is not altered by changing the assumption of the characteristic velocity dispersion of the 
group with planets; there is no significant difference in adopting a Gaussian with the $\sigma$ 
of the whole comparison sample, or of the metal-rich subsample. 

If the kinematic parameters of dynamical streams by \citet{Famaeyetal2005a} are adopted 
(see Table~\ref{test1}, {\it Definition 2}), the results are similar to those reported 
above. The differences between the excesses for the comparison groups obtained  using the 
two kinematic parameter sets are mostly irrelevant. For planet-host stars, the two definitions 
give larger discrepancies. In the {\it Definition 2} case, the group of stars with planets
 behaves in the same way as the metal-rich comparison subsample. However, it is worth noting 
that the analysis by \citet{Famaeyetal2005a} involved K and M giants, while the targets of 
the volume-limited CORALIE survey are mostly G and K dwarfs. In that sense, the determination 
of the dynamical stream parameters could be affected by the target spectral type. The
 kinematic parameters of the dynamical streams obtained in our work are thus considered 
more suitable for our analysis. 
   
The results from the ``overdensity'' method are listed in Table~\ref{test2}. The 
{\it Def.\ 1} column refers to the case adopting the kinematic parameters of the streams 
ellipsoids from our work and [Fe/H] = 0.1 as metallicity cutoff, and subtracting the averaged 
field density to the Hyades overdensity. In the same table we also included the same results 
for different cases: \begin{itemize}
\item {\it Def.\ 2}: adopting the kinematics parameters of the stream ellipsoids by \citet{Famaeyetal2005b};
\item {\it Def.\ 1 mod.\ 1}: adopting our kinematic parameters, but without subtracting the averaged field density;
\item {\it Def.\ 1 mod.\ 2}: adopting our kinematic parameters, but adopting [Fe/H] = 0.05 as metallicity 
cutoff between the metal-rich and solar comparison subsamples;
\item {\it Def. 1 $M<1.2M_{\odot}$}: only the targets with $M<1.2M_{\odot}$ were included 
in the analysis in order to check the effect of age.
\end{itemize} 
The errors have been estimated by the bootstrap method, where the estimate of standard 
error is the standard deviation of the bootstrap replications.

The excess values obtained from the ``counting'' method (see Table~\ref{test1}, 
{\it Definition 1}), and the overdensities obtained from the ``overdensity'' 
analysis (see Table~\ref{test2}, {\it Def.1}), and their corresponding uncertainties, 
are represented in Figure~\ref{res} to offer a more immediate representation. The black 
and red triangles represent the results for planet-host stars, corrected with a Gaussian 
distribution with $\sigma$ equal to the velocity dispersion of the whole comparison sample
 and of the metal-rich subsample, respectively.   

The ``overdensity'' analysis (see Table~\ref{test2}, {\it Def.1}) leads to results similar 
to the ``counting'' method (see Figure~\ref{res}). On the one hand, the comparison metal-rich 
subsample shows a more significant overdensity in Hyades with respect to the whole comparison 
sample, agreeing with what reported by previous studies. On the other hand, the behaviour of 
planet-host stars is consistent with what we observe for the metal-rich CORALIE ``single'' 
stars, whereas it does not seem to reproduce the kinematic characteristic of the whole 
comparison sample. In particular, stars with planets present an overdensity in Hyades 
even more evident than the one found in the metal-rich comparison subsample. Planet-host 
overdensity stands out from the value for the comparison sample in 1.5\,$\sigma$. 

No differences emerge if we adopt the stream parameters by \citet{Famaeyetal2005a} 
(see Table~\ref{test2}, {\it Def.2}), or if we do not subtract the average density
 of the field (see Table~\ref{test2}, {\it Def.1 mod.1}). For the case adopting [Fe/H] = 0.05 
as metallicity cutoff (see Table~\ref{test2}, {\it Def.\ 1 mod.\ 2}), there are no significant 
changes in the results, except for the overdensity of the metal-rich comparison subsample, 
which slightly diminishes. In all the cases, the planet-host stars are much denser in the 
Hyades stream with respect to the whole comparison sample, with a significance of 1.5$\sigma$.

If younger stars are excluded by considering only targets with $M<1.2\,M_{\odot}$, the 
results for the group of stars with planets deviate from the previous ones (see Table~\ref{test2}, 
{\it Def.\ 1 $M<1.2\,M_{\odot}$}). In this case, planet-host stars show a kinematic behaviour 
almost indistinguishable from the metal-rich comparison subsample, with an equivalent 
overdensity in the Hyades.
Younger planet-host stars might be thus responsible for the outstanding excess of stars 
with planets with respect to the rest.

This raises a question concerning the possible effects of age on the kinematic conduct of the 
different groups studied. The small number of targets with $M>1.2\,M_{\odot}$ does not 
allow us to study the behaviour of the younger tail independently, so no conclusive argument 
can be added to discard this possibility. All we can say is that, roughly, a higher 
percentage (around 50\%) of younger stars is found among planet-host targets with Hyades 
stream kinematics than among comparison stars of the Hyades stream.

However, the ratio of younger stars in the planet-host group is the same as that found 
in the metal-rich comparison subsample: 29 younger over 115 older metal-rich comparison 
targets, and 11 younger over 50 older planet-host stars. This ascertainment permits us to 
be more confident that the results are not due to a larger percentage of younger dwarfs in 
the planet-host sample than in the other groups. 
The age effect for planet-host targets seems to be strictly related to their membership
to the Hyades stream. Moreover, this seems to confirm that planet-host stars behave similarly 
as the metal-rich population of the solar neighbourhood. It thus represents a further argument 
in favour of a primordial origin of the metal excess found in stars with giant planets. It cannot be discarded that the important clumpiness of young planet-host stars in the Hyades stream might be due to a more efficient planet formation in the primordial Hyades cluster, progenitor of a significant fraction of young members of the Hyades stream.

All our results suggest that planet host stars follow the kinematic conduct of the metal-rich 
comparison subsample, showing an overdensity in the Hyades stream with respect to the whole comparison 
group. Furthermore, the younger planet-host targets might present a more significant clumpiness 
in the Hyades stream that would result in a further overdensity in the Hyades stream, exceeding that
 found in the metal-rich comparison subsample. However, much more targets are needed to 
confirm this latter suggestion.

This supports the ``primordial'' hypothesis, according to which the metal-rich nature of 
planet-host stars would be due to the high metal content of the protoplanetary clouds from
which the systems formed. Moreover, it can also be understood in the scenario proposed 
as a  possible origin of the dynamical streams \citep[see][and references therein]{Famaeyetal2005b}. 
According to this hypothesis, the kinematic members of the Hyades stream would have formed in 
Galactic inner regions, where the intergalactic medium is more metal-rich. Their 
orbits would have been successively
perturbed by spiral waves, causing radial migration towards the solar neighbourhood. 

The fact that \citet{Jamesetal2006} did not find any metal-rich population in a sample of three nearby young stellar forming regions could support the scenario in which most metal-rich field stars come from inner regions in the Galaxy.
Moreover, \citet[][in preparation]{Chiappini2006} found that Galactic chemical evolutionary models, which reproduce most of the abundance trend peculiarities for thin and thick disc populations, cannot explain the existence of the most metal-rich stars of the solar vicinity (at $\sim$\,8\,kpc). However, those objects are consistent if the same models are computed at inner Galactic radii ($\sim$\,4\,kpc).

\citet{Dehnen2000} proposed the Outer Lindblad Resonance (OLR) as the mechanism reproducing 
the bimodality of the local velocity distribution, identified as the Hercules stream. In 
this model, closed orbits inside the OLR would move slightly outwards, while those outside 
the OLR would move inwards, producing two stellar streams. Since the simulations show that 
the radius of the OLR of the Galactic bar lies inside the solar circle, the stellar velocity 
distribution observed for late-type stars in the solar neighborhood would be affected by the OLR.
These results were then confirmed by further simulations by \citet{Muhlbauer&Dehnen2003}. 
This has been the first example of a non-axysimmetric origin for a stream and has opened up new 
possibilities for relating the other streams with other non-axysimmetric perturbations.

Several simulations have studied the effects of non-axysimmetric perturbations, in particular 
of transient spiral waves, on the galactic disc \citep[e.g.][]{Sellwood&Binney2002,
 Bissantzetal2003,Chakrabarty2004, Desimoneetal2004}. They found that spiral waves can cause 
radial migration near the corotation radius. According to Sellwood \& Binney's model, stars 
just inside corotation would swap places with those outside, and both groups just would exchange 
places. In this scenario, \citet{Famaeyetal2005a} interpreted Hyades and Sirius streams as the 
inward- and outward-moving streams of stars on horshoe orbits that cross rotation. This would 
easily explain the seemingly peculiar metallicity of these moving groups. 

Another interesting feature appears if we inspect and compare the density distributions of 
Galactic space velocities for the metal-rich and solar subsamples (see Figure~\ref{MetalPick}). 
In the Hyades ellipsoid, the peaks corresponding to the overdensities of the two groups are 
displaced: the overdensity of the metal-rich subsample (in the {\it left panel}) is shifted to 
lower $U$ values, whereas the solar subsample ({\it right panel}) shows a peak shifted 
towards higher $U$ velocities. It is also worth noting that the Sirius overdensity completely 
disappears when considering only metal-rich targets (see {\it right panel}). This suggests that
 the members of the Sirius stream have solar and sub-solar metallicities. Therefore, it is in
 good agreement with the scenario proposed by \citet{Famaeyetal2005a}, according to which the
 Sirius stream would correspond to the inward-moving counterpart of the Hyades stream. 

In this framework, the larger overdensity of planet-host stars observed in the Hyades stream 
would suggest that at inner galactic radii, in environments more metal-rich than the solar 
neighborhood, systems with giant planets form more easily. This strongly supports the primordial 
origin proposed to explain the metal excess observed in stars harbouring giant planets 
\citep{Santosetal2000, Santosetal2001}. Current models of planet formation by core accretion 
are in agreement with this observational constraint, and find that metal-rich systems favour the 
formation of massive planets \citep[e.g.][]{Ida&Lin2005, Benzetal2006}. The combination of this
 effect with the detection bias towards more massive planets enables models to reproduce the 
observed correlation between stellar metallicity and the likelihood of harbouring planets \citep{Benzetal2006}.

\section{Concluding remarks}

This is the first study carried out on a volume-limited sample of stars with and without 
exoplanets aimed at investigating the comparative kinematics of planet-host stars and 
their possible relation with the dynamical streams of the solar neighbourhood. With this 
goal, we have compared the group of stars harbouring planets with the rest of the CORALIE 
sample, distinguishing  subgroups of metallicity and age. This was made possible by the precise 
radial velocity measurements and the CCF parameters included in the CORALIE database, and 
using the HIPPARCOS and Tycho-2 catalogues. For targets with detailed abundance determinations, 
a complete study on whether they belong to the thick and thin disc populations, and on possible 
signatures in the abundance trends, has been carried out. The analyses related to the dynamical 
streams have been carried out using two independent methods that have produced equivalent results.

The main outcomes of this work are the following: 
\begin{itemize}
\item The CORALIE planet-host target \object{HD\,4308} has been found to be a very probable
 member of the Galactic thick disc, together with another 21 comparison stars; the dependence 
of the abundance trends on the thin/thick disc populations has been studied for a subset of 
the CORALIE sample with homogeneous abundance determinations from previous studies.
\item No significant differences or peculiar enrichment in $\alpha$-elements are observed 
in the abundance patterns of the subset of stars with homogeneous abundance determinations from 
previous studies. However, the small number of thick disc members prevents us from reaching conclusive 
results. 
\item The tentative existence of thick disc stars at supersolar metallicities is confirmed by 
the appearance of one thick member over 21 with high metallicity (\object{HD\,152391} with [Fe/H] = 0.03). 
\item The high average metallicity of the Hyades stream reported by previous works is confirmed. 
Our results are in agreement with the scenario of non-axysimmetric perturbations, such as transient 
spiral waves, proposed by several results of simulations and observations. According to this model, 
the Hyades stream would be the outward-moving stream of high-metallicity stars born at inner Galactic 
radii and then pushed into the solar neighborhood. We have also found a new clue confirming the solar and subsolar average 
metallicity suggested for the Sirius stream.
\item The group of planet-host stars shows a kinematic behaviour much more similar to the metal-rich 
comparison subsample, than to the whole comparison sample, in the sense that the overdensity in the 
Hyades stream observed for stars with planets is much higher than that observed for the whole 
comparison group. This issue has been reached by the two methods independently and strongly supports 
a primordial origin for the metal excess observed in stars with giant planets. 
\item If we interpret our results within the scenario proposed to explain the presence and origin of 
dynamical streams in the solar neighbourhood, stars with giant planets could have formed more easily 
at inner Galactic radii in a more metal-rich intergalactic medium and then suffered radial 
displacements due to a non-axysimmetric component of the galactic potential, pushing them into 
the solar neighbourhood.
\end{itemize}

This paper presents the results of the first complete analysis on this topic. However, the 
uncertainties affecting our results are large, due to the intrinsic errors of the kinematics 
analysis, but also, and mainly, to the inevitably small number of targets with planets in the 
CORALIE sample. Consequently, substantial improvements will be possible in the near future, 
when the steadily increasing number of known stars with planetary companions will have risen 
enough to permit more definite conclusions. 
\acknowledgement{This research has made use of the SIMBAD database, operated at CDS, Strasbourg,
 France. We thank C. Carretero for many fruitful discussions and useful suggestions. Support from 
Funda\c{c}\~ao para a Ci\^encia e a Tecnologia (Portugal) to N.C.S. in the form of a scholarship 
(reference SFRH/BPD/8116/2002) and a grant (reference POCI/CTE-AST/56453/2004) is gratefully acknowledged.}
 
\bibliographystyle{/home/alex/tex/aa}
\bibliography{aamnem99,aabib}

\end{document}